# Reliably Estimating Bare Chi from Compressible Blends in the Grand Canonical Ensemble

**Sagar S. Rane* and P. D. Gujrati*†**

**Department of Polymer Science,* Department of Physics,†**

**The University of Akron, Akron, OH 44325**

## Abstract

The bare chi characterizing polymer blends plays a significant role in their macroscopic description. Therefore, its experimental determination, especially from small-angle-neutron-scattering experiments on isotopic blends, is of prime importance in thermodynamic investigations. Experimentally extracted quantity, commonly known as the effective chi is affected by thermodynamics, in particular by polymer connectivity, and composition and density fluctuations. The present work is primarily concerned with studying four possible effective chi's, one of which is closely related to the conventionally defined effective chi, to see which one plays the role of a reliable estimator of the bare chi. We show that the conventionally extracted effective chi is not a good measure of the bare chi in *most* blends. A related quantity that does not contain any density fluctuations, and one which can be easily extracted, is a good estimator of the bare chi in all blends except weakly interacting asymmetric blends (see text for definition). The density fluctuation contribution is given by $(\Delta\bar{v})^2/2TK_T$, where $\Delta\bar{v}$ is the difference of the partial monomer volumes and $K_T$ is the compressibility. Our effective chi's are theory-independent. From our calculations and by explicitly treating experimental data, we show that the effective chi's, as defined here, have weak composition dependence and do not diverge in the composition wings. We elucidate the impact of compressibility and interactions on the behavior of the effective chi's and their relationship with the bare chi.

## I. Introduction

A lattice model of an incompressible polymer mixture of species 1 and 2 is characterized by a dimensionless bare exchange energy parameter $\chi_{12} = q\beta\varepsilon_{12}$. Here $q$ is the coordination number of the lattice, $\beta$ is the inverse temperature in the units of the Boltzmann constant, $\varepsilon_{ij} = e_{ij} - 1/2(e_{ii} + e_{jj})$ is the microscopic exchange interaction energy, and $e_{ij}$ the bare or van der Waals interaction energy between species $i$ and $j$. The bare exchange energy appears as a parameter in the partition function of the model, where it determines the energy of various configurations in the system. As such, $\varepsilon_{ij}$ must be independent of the thermodynamic state, i.e. of composition, molecular weight, free volume etc.[1,2] Being a fundamental microscopic parameter, $\chi_{12}$ must remain the same in a homopolymer blend, a block copolymer, etc at the same temperature. Its value, therefore, is of utmost importance in polymer thermodynamics and a great deal of effort has been made to measure it.[3-11] However, what one measures from experiments is not the bare chi, but an effective chi that has been *modified* by thermodynamics. Because of this, it depends on the thermodynamic state of the system. In particular, it develops a composition-dependence, which can be extremely strong in the composition wings if not carefully defined.[1]

In the simplest theory, known as the Flory-Huggins (F-H) theory,[12-14] of an incompressible lattice model of polymers, we find that the effective chi is equal to $\chi_{12}$ due to the random-mixing approximation (RMA).[15] (Here, F-H theory will always refer to the above theory of the incompressible model, and not to any of its possible extensions to the compressible model.) The equality is not true in general. Going beyond the RMA,[16-20] we find that the effective chi has a weak composition-dependence even in the absence of free volume, with its magnitude close to

$$\chi_{\mathrm{NR}} \equiv (q-2)\chi_{12}/q, \tag{1}$$

for polymeric fluids with weak interactions, and small amount of free volume. The prefactor $(q-2)/q$ in the definition of $\chi_{\mathrm{NR}}$ has its origin in the non-randomness (NR) caused by polymer *connectivity* (we ignore end-group effects) and is important for finite $q$ ($\lesssim$12) expected for real systems. Only when $q \to \infty$, the limit in which the RMA becomes valid,[15] do we expect the connectivity to play no significant role and we expect the effective chi in the incompressible model to be identical to $\chi_{12}$. We, thus, conclude that the composition fluctuations and the polymer connectivity have no effect on the effective chi in the incompressible RMA limit (the F-H theory). To see their effects, one must either consider non-random (NR) theories, or real systems that will always exhibit non-randomness.

In an incompressible system, the effective chi is conventionally defined by the following quantity $\Gamma$ related to the second derivative of the Helmholtz free energy $F$ per site: $\Gamma \equiv (1/2)(\partial^2 \beta F/\partial \phi_{\mathrm{m1}}{}^2)$, where $\phi_{\mathrm{m1}}, \phi_{\mathrm{m2}}$ denote the densities of the two species 1 and 2, with $\phi_{\mathrm{m2}} = 1 - \phi_{\mathrm{m1}}$. One then subtracts $\Gamma$ from a reference value

$$\Gamma_{\mathrm{FH,ath}} \equiv (1/2)[1/M_1\phi_{\mathrm{m1}} + 1/M_2\phi_{\mathrm{m2}}] \tag{2}$$

to define the effective chi, known as the F-H chi:

$$\chi_{\rm FH} = \Gamma_{\rm FH,ath} - \Gamma. \tag{3}$$

Using this *subtraction scheme*,[1,4-11,21-28] we easily find that $\chi_{\rm FH} = \chi_{12}$ in the F-H theory. However, as shown previously (see Eq. 32 and discussion following it in Ref. 1), this is not true when the compressibility is nonzero, i.e. when the free volume (represented by voids[29]) is *not* zero. In particular, there emerges a divergence in the composition wings. Similarly, if the experimental data is analyzed using Eq. (3), then $\chi_{\rm FH}$ invariably exhibits a similar divergence. Such a divergence implies that the effective chi has lost its significance as a measure of $\chi_{12}$. In order to avoid this divergence, it was suggested in Ref. 1 that we use in the subtraction scheme an appropriate reference $\Gamma_{\rm ref}$ so as to cancel this divergence. It was shown that using the athermal value $\Gamma_{\rm ath}$ of $\Gamma$ for $\Gamma_{\rm ref}$ ensures the *absence of a divergence* in the wings, regardless of whether $\Gamma$ is calculated in some specific theory or extracted by experiments. Thus, the proposal of Ref. 1 for an effective chi is a *general proposal* applicable to any theory or to any experimental extraction procedure as far as the absence of the divergence is concerned.

The bare chi is required for any first principle calculation or simulation.[30] Only the bare chi can truly characterize the strength and nature of interaction between monomers of two species, regardless of whether we consider a blend or a block copolymer. An effective chi, being dependent on the thermodynamic state, need not be the same in the two systems. Therefore, obtaining a reliable estimator of the bare chi is of utmost importance, and is the central goal of our work. We recall that the aim in Ref. 1 was to demonstrate how to obtain the effective chi without any spurious divergence.

The fluctuations that occur in the system control the quantity $\Gamma$. There are composition fluctuations, but no density fluctuations in an incompressible system. In contrast, a compressible system possesses both kinds of fluctuations. The deviation of $\chi_{\mathrm{FH}}$ from $\chi_{12}$ for a compressible system is caused not only by the presence of both fluctuations, but also due to non-randomness [see Eq. (1)]. Since we have no control over the corrections due to non-randomness, we will only investigate the relative contributions due to the two fluctuations in this paper. There exists an unrealistic ensemble,[1] called the A-ensemble (see the next section for relevant details), in which the free volume and the total volume are kept fixed. Thus, there are only composition fluctuations, but no density fluctuations even though the free volume is not zero. This makes the A-ensemble somewhat similar to the incompressible system as far as the fluctuations are concerned. In contrast, the experiments are done in the grand canonical ensemble, called the C-ensemble,[1] which allows for both fluctuations. Compressibility effects in the A-ensemble are due to composition fluctuations alone and are *minimal*, but by no means absent. In the C-ensemble, both fluctuations determine the compressibility effects. The two ensembles, however, are closely related as demonstrated in Ref. 1, but the relationship was never exploited there. Here, we will exploit this relationship to obtain, within the C-ensemble, a quantity $\Gamma$, which does not contain any contributions from the density fluctuations. The fluctuations are usually measured by performing scattering experiments. The most widely used experimental technique uses small-angle-neutron scattering (SANS) on mixtures of deuterated and hydrogenated polymers for obtaining the effective chi.[31] To emphasize that we are only considering scattering experiments in this work, we will use the subscript 'scatt' in the rest of the paper to represent the effective chi.

The effect of compressibility can be completely documented by dividing blends into two classes: (i) *symmetric* blends in which the two species have equal degrees of polymerization and identical pure-component interactions, and (ii) *asymmetric* blends that do not satisfy either or both of the above conditions. [1] As we will show here, we also find it convenient to classify blends as weakly interacting ($\chi_{12} \lesssim 10^{-4}$–$10^{-5}$, i.e. $\cong 0$), or strongly interacting ($\chi_{12} \gtrsim 10^{-3}$) blends. These numerical values will be justified later.

The primary aim of our theoretical investigation is to identify *the existence of an effective chi that appears to be a reliable estimator of the bare chi.* What remains finally is the important issue concerning its experimental determination with present techniques. We find that the best theoretical choice can be obtained experimentally, except in weakly interacting asymmetric blends. However, it is hoped that our investigation will prompt experimentalists to develop methods to extract the best theoretical choice in the latter case.

The outline of this paper is as follows. In the next section we describe the model and the two ensembles A and C. The two ensembles are further discussed in Sec. III, where we also introduce appropriate $\Gamma$'s, and exploit the relationship between the two ensembles to relate the two $\Gamma$'s. We also introduce two effective chi's in the two ensembles and show their general thermodynamic relationships. The effective chi's do not depend on any particular theory. In Sec. IV, we analyze closely our effective chi's in the RMA limit. In Sec. V, we provide a short discussion of the choice of parameters for our later numerical investigations. The following section contains numerical results for temperatures well above the critical temperature for phase separation. In Sec. VII, we study the behavior of the effective chi's at the critical point and compare them with the results from the previous section. In Sec. VIII, we test our subtraction procedure with SANS data and show that the properly defined effective chi has weak composition dependence. The final section contains the conclusions and a brief summary of our observations. Some of the important results of our chi's are summarized in Table I. Table II contains the results of our analysis of experimental data carried out in Sect. VIII.

## II. Model

We briefly describe the lattice model of the mixture. We refer the reader to Refs. 1, 19, and 32 for details. A *compressible* binary system is a pseudo "ternary" system in which $j$=0 represents voids, and $j$=1,2 the two monodisperse polymeric species of degree of polymerization (DP) $M_j$, respectively. Each monomer of species 1 and 2 or a void occupies one lattice site. We also assume that the volume of the lattice site $v_0$ is a *constant*, irrespective of which species occupies the site. The lattice volume is $V \equiv Nv_0$, where $N$ is the total number of lattice sites. We set $v_0$ = 1 for convenience. With this choice, every volume in the theory is the volume divided by $v_0$. This reminder will be helpful when we need to reinsert $v_0$ in the formulas. If there is any confusion in any definition, we will explicitly exhibit $v_0$.

The chemical potential per monomer is $\mu_{\mathrm{m}j}$, the number of monomers $N_{\mathrm{m}j}$, and the monomer density $\phi_{\mathrm{m}j} \equiv N_{\mathrm{m}j}/N$, ($N \to \infty$, which is implicitly assumed in the following) for $j$=1,2. The number of voids is $N_0$, the void density $\phi_0 = N_0/N$, the total number of monomers $N_{\mathrm{m}}$, and the total monomer density $\phi_{\mathrm{m}} = N_{\mathrm{m}}/N$. We introduce $K_{\mathrm{m}j} = \exp(\beta\mu_{\mathrm{m}j})$. The sum rules are:

$$N \equiv N_0 + N_{\mathrm{m}1} + N_{\mathrm{m}2}; \qquad \phi_0 + \phi_{\mathrm{m}1} + \phi_{\mathrm{m}2} \equiv 1.$$

Corresponding to the bare exchange energies $\varepsilon_{ij}$, $i \neq j$=0,1,2, are the Boltzmann weights $w_{ij}$ =$\exp(-\beta\varepsilon_{ij})$. Let $N_{ij}$ denote the nearest-neighbor unbonded ($i$,$j$) contacts between monomers (including voids) of species $i$ and $j$. The corresponding densities are $\phi_{ij} \equiv N_{ij}/N$.

**A-ensemble.** The A-ensemble is the $N$-$N_0$-$T$-$\mu_m$ ensemble in which the number $N_0$ of voids is kept fixed. Using the activity $K_m = \exp(\beta\mu_m)$, we can represent the corresponding partition function by

$$Z_A(N, N_0, T, \mu_m) \equiv \sum \Omega K_m^{N_{m1}} w_{12}^{N_{12}} w_{01}^{N_{01}} w_{02}^{N_{02}} . \qquad (4a)$$

It can be shown that $K_m$ is related to the activities in the C-ensemble by $K_m = K_{m1}/K_{m2}$ (see below), so that $\mu_m \equiv \mu_{m1} - \mu_{m2}$.[1]

**C-ensemble.** The experimental situation requires considering scattering from a region of *fixed* volume, but with fluctuating $N_{mj}$, their average being controlled by their respective chemical potential $\mu_{mj}$. We therefore consider the C-ensemble,[1] i.e., the $N$-$T$-$\mu_{m1}$-$\mu_{m2}$ ensemble for the compressible system, for which the partition function is given by

$$Z_C(N, T, \mu_{m1}, \mu_{m2}) \equiv \sum \Omega K_{m1}^{N_{m1}} K_{m2}^{N_{m2}} w_{12}^{N_{12}} w_{01}^{N_{01}} w_{02}^{N_{02}} , \qquad (4b)$$

where $\Omega(N, N_{m1}, N_{m2}, N_{12}, N_{01}, N_{02})$ is the number of distinct configurations of polymers. The sum is over all possible *distinct* values of $N_{m1}$, $N_{m2}$, $N_{01}$, $N_{02}$ and $N_{12}$. The free energy per site $\omega_C \equiv (1/N)\ln Z_C$ is the reduced pressure[19] $z_0 \equiv \beta P v_0$.

## III. General Considerations

Let $S_{ij} \equiv (1/Nv_0)\langle\Delta N_{mi}\Delta N_{mj}\rangle$ denote the structure factor, which has the dimension $[S_{ij}]$ of inverse volume. Here, $\Delta N_{mj}$ represents the fluctuation in the quantity $N_{mj}$. The sum $S_{11} + 2S_{12} + S_{22}$ gives the fluctuations in the total number of monomer $N_m = N_{m1} + N_{m2}$. The forward scattering intensity per unit volume is given by

$$I(0) \equiv \sum_{i,j=1,2} b_i b_j S_{ij} ,$$

where $b_j$ is the coherent scattering length for species $j$. The dimension of $I(0)$ is 1/length.

**1. A-ensemble** We briefly review this ensemble, which has been extensively studied in Ref. 1; we refer the reader to this work for full details. The ensemble also describes the *incompressible* model. There is *no* density fluctuation, and $\Delta N_{m1} = -\Delta N_{m2}$. Thus, $S_{11}^{(A)} = S_{22}^{(A)} = -S_{12}^{(A)}$, and

$$I(0) = (b_1 - b_2)^2 S_{11}^{(A)}; \tag{5}$$

the intensity is governed only by the composition fluctuation. Dividing $I(0)$ by $k_N \equiv (b_1 - b_2)^2/v_0 = (b_1 - b_2)^2$, we obtain a *purely thermodynamic* and *dimensionless quantity* $\Gamma_A$:

$$2\Gamma_A \equiv k_N/I(0) = 1/S_{11}^{(A)},$$

and, as shown in Ref. 1, it is related to the derivative of the chemical potential difference:

$$2\Gamma_A(\chi_{12}, \chi_{01}, \chi_{02}) \equiv (\partial\Delta\beta\mu/\partial y)_{T,\phi_0}/(1-\phi_0), \tag{6}$$

where $\Delta\mu \equiv \mu_{m2} - \mu_{m1}$.

**2. C-ensemble** The structure factors are $S_{ij}^{(C)} \equiv (1/Nv_0)(\partial N_{mi}/\partial\beta\mu_{mj})_{V,T,\mu'}$; $\mu'$ denotes the remaining $\mu_{mj}$ not used in differentiation. Introducing

$$C_{12} \equiv -(1/N)(\partial\beta\mu_{m1}/\partial N_{m2})^{-1}_{T,P,N_{m1}},$$

$$\overline{b} = (b_1\phi_{m1} + b_2\phi_{m2}), \quad \Delta\widetilde{b} = (b_1\overline{v}_2 - b_2\overline{v}_1), \tag{7}$$

we have[33]:

$$S_{11}^{(C)} = TK_T\phi_{m1}^2 + \phi_{m1}\phi_{m2}\overline{v}_2^2 C_{12}, \; S_{22}^{(C)} = TK_T\phi_{m2}^2 + \phi_{m1}\phi_{m2}\overline{v}_1^2 C_{12},$$

$$S_{12}^{(C)} = TK_T\phi_{m1}\phi_{m2} - \phi_{m1}\phi_{m2}\overline{v}_1\overline{v}_2 C_{12}. \tag{8}$$

The intensity $I(0)$ consists of two terms,[33]

$$I(0) \equiv I_1 + I_2, \;\; I_1 \equiv \overline{b}^2 TK_T, \; I_2 \equiv \phi_{m1}\phi_{m2}\Delta\widetilde{b}^2 C_{12}, \tag{9}$$

where $\bar{v}_1, \bar{v}_2$ are the *partial* monomer volumes of species 1 and 2 respectively and $K_T$ the isothermal compressibility at fixed monomer numbers. The intensity depends on extraneous quantities $b_1$, and $b_2$, which cannot be removed from $I(0)$. Thus, $I(0)$ cannot be used to obtain a purely thermodynamic quantity.[31] Note that both parts of the intensity have an implicit $1/v_0$ factor for dimensional reasons.

It is easy to show that

$$C_{12}^{-1} = (N_{\mathrm{m}1} N_{\mathrm{m}2} N / N_{\mathrm{m}}^3)(\partial^2 \beta g_{\mathrm{m}} / \partial y^2)_{T,P}\,, \tag{10}$$

$$(\partial \beta g_{\mathrm{m}} / \partial y)_{T,P} = \Delta\beta\mu \equiv \beta\mu_{\mathrm{m}2} - \beta\mu_{\mathrm{m}1}\,. \tag{11}$$

where $y \equiv \bar{\phi}_{\mathrm{m}2} \equiv \phi_{\mathrm{m}2}/(1-\phi_0)$, $\phi_{\mathrm{m}1} \equiv 1-y$, and where $g_{\mathrm{m}} \equiv G/N_{\mathrm{m}}$ is the Gibbs free energy $G$ per monomer. We observe that $\Delta\tilde{b}^2/I_2$ is a purely thermodynamic quantity, and is related to the second derivative of the free energy, or the first derivative of $\Delta\mu$; compare with Eq. (6). Thus, we introduce a *purely thermodynamic* and *dimensionless quantity* $\Gamma$ in this ensemble as below:

$$\begin{aligned} 2\Gamma_{\mathrm{C}}(\chi_{12},\chi_{01},\chi_{02}) &\equiv k_{\mathrm{N}}'/I_2 \equiv (\partial^2 \beta g_{\mathrm{m}}/\partial y^2)_{T,P}/(1-\phi_0), \\ &\equiv [C_{12} y(1-y)]^{-1} = (\partial \Delta\beta\mu/\partial y)_{T,P}/(1-\phi_0), \end{aligned} \tag{12}$$

which can be calculated in *any* theory. Here we have used Eqs. (9-11) to re-express $\Gamma_{\mathrm{C}}$. We have also introduced a new contrast factor of dimension 1/length:

$$k_{\mathrm{N}}' \equiv \Delta\tilde{b}^2(1-\phi_0)^2/v_0.$$

We also choose this definition of $\Gamma_{\mathrm{C}}$ because it gives an effective chi that is almost equal to the bare $\chi_{12}$ in the RMA limit; as we show in the next section, see Eq. (23).

Using the subtraction scheme similar to that in Eq. (3), we now introduce our effective chi's in each of the two ensembles. The most commonly used reference state is the athermal reference state (ARS), in which $\Gamma_{\alpha,\mathrm{ref}} = \Gamma_{\alpha,\mathrm{ath}} \equiv \Gamma_\alpha(0,0,0)$, ($\alpha$ = A or C) is the *athermal* part of $\Gamma_\alpha$, and is trivially obtained by considering the system at infinite temperatures. But it is important to note that in evaluating $\Gamma_\alpha(0,0,0)$ the densities $\phi_{\rm m1}$ and $\phi_{\rm m2}$ are held fixed. Thus, the difference $\Gamma_{\alpha,\rm int} \equiv \Gamma_\alpha - \Gamma_{\alpha,\rm ath}$ known as the interaction part of $\Gamma_\alpha$ is equal to the negative of $\chi_{\rm scatt}^{(\alpha)}$ with this choice of the reference state. Due to the choice of the ARS, we see that $\chi_{\rm scatt}^{(\alpha)}$ is zero only when *all* three bare chi's are zero, and not when just $\chi_{12}$=0.

An effective chi that vanishes when *only* $\chi_{12}$ vanishes is obviously a better measure of $\chi_{12}$. Therefore, using the interacting reference state (IRS) in which $\chi_{12}$=0, but $\chi_{01}$ and $\chi_{02}$ have the values appropriate for the pure components and denoting the corresponding $\Gamma_{\alpha,\rm ref}$ by $\overline{\Gamma}_{\alpha,\rm ref} \equiv \Gamma_\alpha(0,\chi_{01},\chi_{02})$, another effective chi viz. $\overline{\chi}_{\rm scatt}^{(\alpha)}$ is similarly obtained. Again, in evaluating $\overline{\Gamma}_{\alpha,\rm ref}$, the densities are held fixed to the values at which $\Gamma_\alpha$ is evaluated.

Our two effective chi's are

$$\begin{aligned}\chi_{\rm scatt}^{(\alpha)} &\equiv \Gamma_\alpha(0,0,0) - \Gamma_\alpha(\chi_{12},\chi_{01},\chi_{02}),\\ \overline{\chi}_{\rm scatt}^{(\alpha)} &\equiv \Gamma_\alpha(0,\chi_{01},\chi_{02}) - \Gamma_\alpha(\chi_{12},\chi_{01},\chi_{02}).\end{aligned} \qquad (\alpha = {\rm A,C}). \tag{13}$$

In terms of $\chi_{\rm scatt}^{(\alpha)}$, we have

$$\overline{\chi}_{\rm scatt}^{(\alpha)} \equiv \chi_{\rm scatt}^{(\alpha)}(\chi_{12},\chi_{01},\chi_{02}) - \chi_{\rm scatt}^{(\alpha)}(0,\chi_{01},\chi_{02}), \quad (\alpha = \text{A or C}). \tag{14}$$

**3. Extracting $\chi_{\rm scatt}^{(\rm A)}$ and $\overline{\chi}_{\rm scatt}^{(\rm A)}$ in the C-ensemble** Consider the following important identity due to Gujrati[1]

$$S_{11}^{(\rm A)} = \Delta_{\rm C}/(S_{11}^{(\rm C)} + 2S_{12}^{(\rm C)} + S_{22}^{(\rm C)}), \tag{15}$$

where $\Delta_C \equiv \det S^{(C)} = S_{11}^{(C)} S_{22}^{(C)} - (S_{12}^{(C)})^2$. From Eqs. (6, 8, 12, and 15), we obtain,

$$2\Gamma_A = 2\Gamma_C + (\Delta\bar{v})^2 / TK_T, \tag{16}$$

where we have introduced $\Delta\bar{v} \equiv \bar{v}_1 - \bar{v}_2$. The above relation is important since it relates the $\Gamma$ in the two ensembles. From the property of the A-ensemble, we know that $\Gamma_A$ contains contributions from the composition fluctuations alone, while $\Gamma_C$ contains the additional contribution from the density fluctuations.[34,35] Hence, one-half of the last term in Eq. (16) represents the *density-fluctuation contributions* to $\Gamma$. This contribution can be easily measured by present experimental techniques. To achieve this, experimentalists need $K_T$, $\bar{v}_i$, and $\phi_0$. The free volume is given by

$$\phi_0 = 1 - [(1-y)\bar{v}_1 + y\bar{v}_2]^{-1}. \tag{17}$$

Hence, Eq. (16) allows us to obtain the A-ensemble quantity from measurements carried out in the C-ensemble. Since the A-ensemble exhibits minimal compressibility effects and there are no density fluctuations,[36] $\chi_{\text{scatt}}^{(A)}$ turns out to be a *better* estimator of $\chi_{12}$, Upon subtracting out the athermal part from both sides in Eq. (17) we have,

$$\chi_{\text{scatt}}^{(C)} = \chi_{\text{scatt}}^{(A)} + \frac{1}{2T}\left[(\Delta\bar{v})^2 / K_T - (\Delta\bar{v}_{\text{ath}})^2 / K_{T,\text{ath}}\right]. \tag{18}$$

It is obvious that $\bar{v}_{1,\text{ath}} = \bar{v}_{2,\text{ath}}$ for equal DP's. Therefore,

$$\chi_{\text{scatt}}^{(C)} = \chi_{\text{scatt}}^{(A)} + (\Delta\bar{v})^2 / 2TK_T, \ (M_1{=}M_2). \tag{19}$$

Eqs. (18, 19) allow us to calculate $\chi_{\text{scatt}}^{(A)}$ associated with the *unphysical* A-ensemble from the measurement of $\chi_{\text{scatt}}^{(C)}$. To obtain $\chi_{\text{scatt}}^{(C)}$, experimentalists need to extract $\Gamma_C$. The last term in Eq. (16) vanishes as $\phi_0 \to 0$ or 1 as it must, but for two different reasons. In the incompressible limit, the numerator vanishes faster than the denominator. In the gas phase, the denominator diverges. Therefore, the last term in Eq. (18) captures the *additional* compressibility contributions, not included in $\chi_{\mathrm{scatt}}^{(\mathrm{A})}$.

From the thermodynamic identity

$$(\partial\phi_0/\partial y)_{T,P} = -\phi_{\mathrm{m}}^2\Delta\overline{v}\,, \tag{20}$$

which is easily proven, we observe that $\Delta\overline{v}$ is usually not zero, since the free volume usually changes with *y*. Thus, one cannot neglect the difference in the partial monomer volume. Neglecting this difference is equivalent to treating the free volume as constant, which is what is expected in the A-ensemble. This explains the relationship between the two ensembles as it appears in Eqs. (18, 19), in which the last term on the right-hand side describes the effect of free-volume variation.

## IV. RMA Limit of the Effective Chi

We now consider the RMA limit of our theory.[16-19] The Helmholtz free energy *F* per unit volume[18] ($f_i$ is the embedding constant of the *i*-species polymer) in this limit is given by:

$$\begin{aligned}-\beta F \;\to\; &(\phi_{\mathrm{m1}}/M_1)\ln\left(f_1/\phi_{\mathrm{m1}}\right)+(\phi_{\mathrm{m2}}/M_2)\ln\left(f_2/\phi_{\mathrm{m2}}\right)\\ &-\phi_0\ln\phi_0-\chi_{12}\phi_{\mathrm{m1}}\phi_{\mathrm{m2}}-\chi_{01}\phi_0\phi_{\mathrm{m1}}-\chi_{02}\phi_0\phi_{\mathrm{m2}}\end{aligned}\;; \tag{21}$$

terms containing $f_i$ (which formally diverge as $q\to\infty$) do not contribute when we calculate the second derivative of the Helmholtz free energy. Introducing $\Delta\chi_0=\chi_{01}-\chi_{02}$ and using Eq. (11), we find in the RMA limit,

$$(\partial\beta g_{\mathrm{m}}/\partial y)_{T,P}\to(1/M_2)\ln\phi_{\mathrm{m2}}-(1/M_1)\ln\phi_{\mathrm{m1}}+\chi_{12}(\phi_{\mathrm{m1}}-\phi_{\mathrm{m2}})-\Delta\chi_0\phi_0,$$

from which we easily find that ( $\delta M'\equiv 1/M_1-1/M_2$ )

$$(\partial^2 \beta g_{\rm m} / \partial y^2)_{T,P} \to (1/M_1\phi_{\rm m1} + 1/M_2\phi_{\rm m2})(1-\phi_0) - 2\chi_{12}(1-\phi_0) \\ + [\delta M'/(1-\phi_0) - \chi_{12}(1-2y) - \Delta\chi_0](\partial\phi_0/\partial y)_{T,P} , \quad (22)$$

in the RMA limit. Using Eqs. (12), (13), (20), and (22), we find that

$$\chi_{\rm scatt}^{\rm (C)} \to \chi_{12} - \phi_{\rm m}\Delta\bar{v}[\chi_{12}(1-2y) + \Delta\chi_0]/2 + \delta M'(\Delta\bar{v} - \Delta\bar{v}_{\rm ath})/2 \ . \quad (23)$$

The last two terms clearly show that there remains some contribution from the compressibility to $\chi_{\rm scatt}^{\rm (C)}$, even in the RMA limit. In addition, it also depends on the difference $\Delta\chi_0$. This causes problems for blends in which $\chi_{12}$ is extremely small. In such cases, the residual contribution can become comparable or even large, and can make $\chi_{\rm scatt}^{\rm (C)}$ an unreliable estimator of $\chi_{12}$. This problem for weak blends will remain present in any theory including our theory.

It is important to note that for a symmetric blend, Eq. (23) contains only $\chi_{12}$:

$$\chi_{\rm scatt}^{\rm (C)} \to \chi_{12}[1 - \phi_{\rm m}\Delta\bar{v}(1-2y)/2] . \quad (24)$$

However, the composition and compressibility effects are still present. Only if $\Delta\bar{v}$ vanishes, as will be the case for the incompressible limit, or at equal composition ($y = 1/2$), can we expect the effective chi and the bare chi to be the same in the RMA limit. We should contrast $\chi_{\rm scatt}^{\rm (C)}$ in Eq. (23) with the RMA limit of $\chi_{\rm scatt}^{\rm (A)}$ in the A-ensemble:

$$\chi_{\rm scatt}^{\rm (A)} \to \chi_{12} ; \quad (25)$$

see Eq. (58) in Ref 1. It is evident that $\chi_{\rm scatt}^{\rm (A)}$ is a better estimator of $\chi_{12}$ than $\chi_{\rm scatt}^{\rm (C)}$. (Recall that in the RMA limit, $\chi_{\rm NR} \to \chi_{12}$.) Thus, the effective chi in this limit does not account for polymer connectivity, as expected from our discussion immediately following Eq. (1). We also find that in the RMA limit,

$$\overline{\chi}_{\rm scatt}^{\rm (C)} \to \chi_{12} - [\phi_{\rm m}\Delta\overline{v}\chi_{12}(1-2y) + (\phi_{\rm m}\Delta\chi_0 - \delta M')(\Delta\overline{v} - \Delta\overline{v}_{\rm IRS})]/2 . \qquad (26)$$

Here, $\Delta\overline{v}_{\rm IRS}$ denotes the value of $\Delta\overline{v}$ in the IRS ($\chi_{12} = 0, \chi_{01}, \chi_{02}$). In the incompressible RMA limit, the effective chi's in Eqs. (23, 24), and (26) all reduce to $\chi_{12}$.

In our recursive lattice theory,[16-19] we can show that $\Gamma_{\rm C,ath} \equiv \Gamma_{\rm C}(0,0,0)$ is given by,

$$\Gamma_{\rm C,ath} = -\frac{(1-\phi_0)^2}{2\phi_{\rm m1}\phi_{\rm m2}\phi_u\phi_0}\left[(\phi_u - \nu_1\nu_2\phi_0) - \frac{(\phi_u - q\nu_1\phi_0/2)(\phi_u - q\nu_2\phi_0/2)}{(1-\phi_0)\phi_u - q\phi\phi_0/2}\right], \qquad (27a)$$

where $\phi_u = q/2-\phi$ denotes the fraction of uncovered lattice bonds, $\phi$ the total bond density, and $\nu_j \equiv 1-1/M_j$. The most dominant part of $\Gamma_{\rm C,ath}$ in the wings is $\Gamma_{\rm FH,ath}$, given in Eq. (2), which is the (most) diverging contributions in the incompressible FH theory, and the compressible RMA theory [see Eq. (22)]. This strongly suggests that *if an appropriate* $\Gamma_{\rm C}$ *is so identified that the leading term in its interaction part reduces to* $\chi_{12}$ *in the RMA limit, then* $\Gamma_{\rm C,ath}$ *will have* $\Gamma_{\rm FH,ath}$ *as its most dominant contribution.* For $M_1$=$M_2$, we find that

$$\Gamma_{\rm C,ath} \equiv \Gamma_{\rm FH,ath} \ (M_1 = M_2). \qquad (27b)$$

## V. Numerical Analysis and Choice of Parameters

We now turn our attention to check the relative values of various effective chi's in comparison with $\chi_{\rm NR}$. For this purpose, we use exclusively our recursive theory,[16-19] which goes beyond the RMA The symmetric case exhibits a weak composition and compressibility dependence in all cases we study. The asymmetric case is capable of exhibiting stronger composition, as well as compressibility dependence. The choice of the parameters is discussed in detail in Sects. V (B), and (C), Ref. 1. For many polymer systems (such as PS, PVME, and PE), the applicable values of $w_{01}$ and $w_{02}$, and of $w_{12}$, are in the range of 0.75-0.85, and 0.9-1.0, respectively. In most of our calculations, we have chosen the ranges in order to describe realistic systems. Architectural differences between monomers and/or deuterium labeling (isotopic blends) can bring about an asymmetry in $w_{01}$ and $w_{02}$, and we consider such blends by choosing $w_{01}$ and $w_{02}$ slightly different from each other ( the difference being $\cong$ 0.0065), but in the vicinity of 0.8. If the isotopic blend is of different species, the asymmetry in $w_{01}$ and $w_{02}$ can be even larger. In all calculations, we set $q$=8. In most cases of asymmetric blends that we investigate here, the last term in Eq. (14) is of the order of $10^{-5}$; see Figs. 6(a, b) for $\chi_{12}$=0 in Ref. 1, and the results to be presented later (Sect. VI). For symmetric blends, this term is even smaller. Thus, its effect is minimal and can be neglected for symmetric blends, and strongly interacting blends for which the first term is much larger. In this case, $\chi^{(\alpha)}_{\text{scatt}}$ $\cong \overline{\chi}^{(\alpha)}_{\text{scatt}}$. We must distinguish between $\chi^{(\alpha)}_{\text{scatt}}$, and $\overline{\chi}^{(\alpha)}_{\text{scatt}}$ if the two terms in Eq. (14) are of the same order as $10^{-5}$, which will occur most often in weakly interacting asymmetric blends. This observation now justifies our earlier classification of blends into weakly interacting ($\chi_{12} \lesssim 10^{-4}$–$10^{-5}$, i.e. $\cong$ 0), and strongly interacting ($\chi_{12} \gtrsim 10^{-3}$) blends.

Our plots for effective chi's as a function of composition are done at specified values of $\phi_0$: we choose only those points in phase space that correspond to the specified value of $\phi_0$. This approach considerably simplifies our calculations. The presence or absence of divergences obtained in effective chi's along the specified $\phi_0$ plane will persist when plotted as a function of $y$ at fixed pressure. Thus, the investigation in the fixed $\phi_0$ plane is completely general.

## VI. Numerical Results ($T$>$T_c$)

As discussed in Sect. V, the bar and unbar chi's turn out to be *almost identical* ($\chi_{\rm scatt}^{(\alpha)} \cong \overline{\chi}_{\rm scatt}^{(\alpha)}$) for symmetric and for strongly interacting blends. Therefore, we only show unbar quantities for them.

### A. Symmetric Blends ($w_0$=$w_{01}$=$w_{02}$, $M_1$=$M_2$)

**1. Effect of $w_0$:** In Fig. 1, we show $\chi_{\rm scatt}^{\rm (C)}$ (filled symbols) and $\chi_{\rm scatt}^{\rm (A)}$ (empty symbols) for $w_{12} = 0.9975$ ($\chi_{12} \cong 0.02$, $\chi_{\rm NR} \cong 0.015$), and for two choices of $w_0$=0.76 (●,○) and 1.1 (▼,∇). We immediately note that $\chi_{\rm scatt}^{\rm (A)} \cong \chi_{\rm scatt}^{\rm (C)}$, this is because the last term in Eq. (19) is extremely small for symmetric blends. They also have almost identical composition dependence. Moreover, their values change only *minimally* even though $w_0$ changes dramatically from a repulsive to an attractive interaction. It appears, thus, that $\chi_{\rm scatt}^{\rm (C)}$ and $\chi_{\rm scatt}^{\rm (A)}$ are relatively insensitive to $w_0$ and are mostly determined by $w_{12}$ for symmetric blends and satisfy $\chi_{\rm scatt}^{\rm (A)} \cong \chi_{\rm scatt}^{\rm (C)} \cong \chi_{\rm NR} \cong 0.015$ for small $\phi_0$. As $w_0$ decreases, the voids are repelled from the vicinity of both polymer species. This increases their mutual contacts; hence, $\chi_{\rm scatt}^{\rm (C)}$ increases.

**2. Effect of $\phi_0$:** We show the effect of free volume in Fig. 2 where we consider four different values of $\phi_0$=0.01 (●), 0.05 (○), 0.1 (▼) and 0.99 (∇). We choose $w_0$=1 for simplicity and set $w_{12}$=0.9975 ($\chi_{12} \cong 0.02$, $\chi_{\rm NR} \cong 0.015$) so that the blend is in a single phase. We find that $\chi_{\text{scatt}}^{(\text{C})}$ decreases with increasing free volume for all *y*. Note, however, that $\chi_{\text{scatt}}^{(\text{C})}$ does not vary much as $\phi_0$ increases from 0.01 to 0.99. It also does not vanish as $\phi_0 \to 1$. These are desirable properties in an effective chi, which, while depending on $\phi_0$, should remain non-zero as $\phi_0 \to 1$.

**3. Effect of DP:** We have observed that the weak composition dependence changes very little with the DP in symmetric blends (results not shown). As the DP decreases, the mutual contacts between the two species increases, and we find a higher $\chi_{\text{scatt}}^{(\text{C})}$, but the usual pattern $\chi_{\text{scatt}}^{(\text{A})} \cong \chi_{\text{scatt}}^{(\text{C})} \cong \chi_{\text{NR}}$ is still valid.

## B. Strongly Interacting Asymmetric Blends

**1. Effect of DP-asymmetry:** The DP-asymmetry $|M_1 - M_2|$ changes the composition dependence slightly, but we still have $\chi_{\text{scatt}}^{(\text{C})} \cong \chi_{\text{scatt}}^{(\text{A})} \cong \chi_{\text{NR}}$. We summarize our findings but show no results. The effective chi's are larger in the wing where the mixture is composed mostly of the pure component of higher DP. The "symmetric" form of $\chi_{\text{scatt}}^{(\text{C})}$ and $\chi_{\text{scatt}}^{(\text{A})}$ becomes highly skewed with the aspect ratio $a \equiv M_1/M_2$. However, our calculations show that the composition dependence becomes insensitive to higher DP-asymmetry for a given ratio *a*.

**2. Effect of asymmetry in $w_{01}$ and $w_{02}$:** The effect of asymmetry $\Delta w = |w_{01} - w_{02}|$ in $w_{01}$ and $w_{02}$, however, is much stronger than the DP asymmetry, see $\chi_{\text{scatt}}^{(\text{A})}$ (□) and $\chi_{\text{scatt}}^{(\text{C})}$ (■) in Fig. 1 for $w_{01} = 0.76$, $w_{02} = 0.79$, and $w_{12} = 0.9975$. We immediately observe that the minimum in $\chi_{\text{scatt}}^{(\text{C})}$ and $\chi_{\text{scatt}}^{(\text{A})}$ that is present in the symmetric blend ($w_0$=0.76) has disappeared. The *y*-dependence is of opposite sign in the two chi's due to the ensemble difference. We observe that $\chi_{\rm scatt}^{(A)} \cong \chi_{\rm NR}$. We also observe that $\chi_{\rm scatt}^{(C)}$, which is very different from $\chi_{\rm NR}$ in magnitude, decreases monotonically (≅5%) with *y*; this drop is due to our choice of $w_{02}>w_{01}$ ($\chi_{02}<\chi_{01}$). As $\Delta w$ increases, the variation in $\chi_{\rm scatt}^{(C)}$ over the composition range increases, and remains much larger than that in $\chi_{\rm scatt}^{(A)}$. Thus, not only is $\chi_{\rm scatt}^{(C)}$ more strongly affected by $\Delta w$ than $\chi_{\rm scatt}^{(A)}$, its magnitude is also much larger than that of $\chi_{\rm scatt}^{(A)}$. The difference is due to the second contribution in Eq. (19), which is now large enough.

### C. Weakly Interacting Asymmetric Blends

The bar and unbar effective chi's are now very different. Therefore, we exhibit both of them. We numerically establish that for these blends the bar effective chi's are good estimators of $\chi_{\rm NR}$; unbar effective chi's are extremely unreliable.

**1. Isotopic blend of same species:** We investigate two weakly interacting asymmetric blends ($w_{12}$=0.999999 $\Rightarrow \chi_{\rm NR} \cong 6{\rm x}10^{-6}$, Fig. 3), which resemble an isotopic blend of the same species as the difference $\Delta w$=0.0065 is extremely small. The system has a ratio of hydrogenated to deuterated energies of 0.97, which is also found[27] in a dPE/hPE blend. The empty symbols are for a blend with equal DP=5000. The filled symbols are for a 100/5000 blend. We first consider the blend with equal DP. We find that $\chi_{\rm scatt}^{(C)}$ (□) overestimates $\chi_{12}$ by a factor of 17, and $\chi_{\rm NR}$ by a factor of 23, while $\chi_{\rm scatt}^{(A)}$ (◇) has the *wrong* sign. The latter result was already seen in Ref. 1. With the DP-asymmetry, we find that $\chi_{\rm scatt}^{(C)}$ (■) still overestimates $\chi_{12}$, but $\chi_{\rm scatt}^{(A)}$ (◆) has the *correct* sign, and is much closer to $\chi_{\rm NR}$ than is $\chi^{\rm (C)}_{\rm scatt}$. The difference between $\chi^{\rm (C)}_{\rm scatt}$ and $\chi^{\rm (A)}_{\rm scatt}$ is due to the last term in Eq. (18, 19), and clearly demonstrates the important contribution due to *nonzero compressibility* in weakly interacting blends. We also show $\overline{\chi}^{\rm (A)}_{\rm scatt}$ (▽,▼) and $\overline{\chi}^{\rm (C)}_{\rm scatt}$ (○,●). They lie on top of each other on the scale of the graph and are equal to $\chi_{\rm NR} \cong 6{\rm x}10^{-6}$. Thus, *both bar chi's provide a very good estimate of* $\chi_{\rm NR}$ *for weakly interacting asymmetric blends*. In contrast, *only* $\overline{\chi}^{\rm (A)}_{\rm scatt} \cong \chi^{\rm (A)}_{\rm scatt}$ *provides a good estimate of* $\chi_{\rm NR}$ for symmetric and strongly interacting asymmetric blends. For weak 1-2 interaction, $\chi^{\rm (C)}_{\rm scatt}$ is strongly influenced by $\Delta w$, and can differ considerably from $\chi_{\rm NR}$.

These results have important implications for effective chi's of isotopic blends of the same species, considered in Fig. 3. We clearly see that even at $y$ = 0.5, $\chi^{\rm (C)}_{\rm scatt}$ overestimates $\chi_{\rm NR}$ by more than one order of magnitude and $\chi^{\rm (A)}_{\rm scatt}$ is negative, even though the system has only repulsive interactions. Thus, in this case, both $\chi^{\rm (C)}_{\rm scatt}$ and $\chi^{\rm (A)}_{\rm scatt}$ fail to estimate $\chi_{\rm NR}$.

**2. Effect of $\phi_0$:** We show $\chi^{\rm (C)}_{\rm scatt}$ and $\chi^{\rm (A)}_{\rm scatt}$ for two blends with $\phi_0$ =0.05 (○,●) and 0.1 (∇,▼) in Fig. 4(a), and compare them with $\chi_{\rm NR} \cong 0.0006$. We find that $\chi^{\rm (C)}_{\rm scatt}$ is greater than $\chi_{\rm NR}$. However, $\chi^{\rm (A)}_{\rm scatt}$ is *close* to $\chi_{\rm NR}$. In Fig. 4(b), we show the bar quantities. *The use of* IRS *has brought the bar quantities in line with* $\chi_{\rm NR}$ in both cases. However, $\overline{\chi}^{\rm (C)}_{\rm scatt}$ and $\overline{\chi}^{\rm (A)}_{\rm scatt}$ have opposite slopes, as was the case with strongly interacting asymmetric blend in Fig. 1.

We thus conclude that for symmetric and strongly interacting asymmetric blends, we obtain a highly reliable estimate of the bare chi by evaluating $\chi_{\text{scatt}}^{(\text{A})}$. If $M_1$=$M_2$, by simply subtracting $(\Delta\bar{v})^2/2TK_T$ from $\chi_{\text{scatt}}^{(\text{C})}$, we can get $\chi_{\text{scatt}}^{(\text{A})}$, otherwise we need to use Eq. (18). Thus, it is feasible experimentally to obtain a reliable estimate of the bare chi in these cases.

Our observation that for weakly interacting blends, $\bar{\chi}_{\text{scatt}}^{(\text{C})} \cong \bar{\chi}_{\text{scatt}}^{(\text{A})}$ can be understood as follows. From Eq. (14) and (18), we get

$$\bar{\chi}_{\text{scatt}}^{(\text{C})} = \bar{\chi}_{\text{scatt}}^{(\text{A})} + \frac{1}{2T}\left[(\Delta\bar{v})^2/K_T - (\Delta\bar{v}_{\text{IRS}})^2/K_{T,\text{IRS}}\right], \tag{28}$$

where $K_{T,\text{IRS}}$ denotes the value of $K_T$ in the IRS. In a weakly interacting blend $(\chi_{12} \cong 0)$, each term in the square bracket is largely dictated by $\chi_{01}$ and $\chi_{02}$, and hence $(\Delta\bar{v})^2/K_T \cong (\Delta\bar{v}_{\text{IRS}})^2/K_{T,\text{IRS}}$, and therefore $\bar{\chi}_{\text{scatt}}^{(\text{C})} \cong \bar{\chi}_{\text{scatt}}^{(\text{A})}$.

### D. Temperature Dependence

The bare chi's have a trivial $1/T$ temperature dependence. In order to characterize any additional temperature-dependence of the effective chi's, we evaluate the ratio $R_\alpha \equiv \chi_{12}/\chi_{\text{scatt}}^{(\alpha)}$, which should exhibit no $T$ dependence if $\chi_{\text{scatt}}^{(\alpha)} \propto 1/T$. In Fig. 5, we plot $R_\text{A}$, and $R_\text{C}$ as a function of $T$ for a 100/100 blend. The lowest temperature we choose corresponds to a critical point ($y$=0.497, $\phi_0$=0.05, $w_{01}$=0.77, $w_{02}$=0.76, $w_{12}$=0.99664) for the blend, which we arbitrarily assign a temperature of 100 (no units) and obtain $R_\text{A}$ and $R_\text{C}$ (at fixed $y$ and $\phi_0$) for higher temperatures, which changes $w_{01}$, $w_{02}$, and $w_{12}$. We find that $\chi_{\text{scatt}}^{(\text{A})}$ has approximate $1/T$-dependence; some deviation from a simple $1/T$ behavior is seen at low temperatures. However, $\chi_{\text{scatt}}^{(\text{C})}$ has a complicated *T*-dependence. This again demonstrates that $\chi_{\text{scatt}}^{(\text{A})}$ behaves more like the bare chi than $\chi_{\text{scatt}}^{(\text{C})}$.

From all the above results we conclude that $\chi_{\text{scatt}}^{(\text{C})}$ *is not a reliable estimator of* $\chi_{\text{NR}}$ *in all cases*. However, $\chi_{\text{scatt}}^{(\text{A})}$ is a good candidate for estimating $\chi_{\text{NR}}$ in symmetric and in strongly interacting asymmetric blends. In weakly interacting asymmetric blends, only $\overline{\chi}_{\text{scatt}}^{(\text{C})}$ and $\overline{\chi}_{\text{scatt}}^{(\text{A})}$ provide the best possible estimate of the bare chi through $\chi_{\text{NR}}$. For the benefit of the reader we summarize our numerical findings in Table I.

## VII. Behavior at $T_c$

We now study the behavior of the effective chi at the critical point $T_c$, as many blends happen to be near their critical point under ordinary conditions. We calculate $\chi_{\text{scatt}}^{(\text{C})}$ and $\chi_{\text{scatt}}^{(\text{A})}$ at the critical point for a number of systems and compare them with the bare $\chi_{12}$. Every state in our model is determined uniquely by five parameters: $\phi_0$, $y$, $w_{01}$, $w_{02}$ and $w_{12}$. The critical point however is determined by two constraints: (i) the determinant $\Delta_{\text{C}}$ diverges ($\Rightarrow \Gamma_{\text{C}}=0$, spinodal condition), and (ii) $T_c$ is the highest temperature on the spinodal. Hence, we have three free parameters, which we choose to be $\phi_0$, $w_{01}$ and $w_{02}=1.56-w_{01}$. We then find unique values of $y$ and $w_{12}$ are at the critical point. We cover the range $0.76 \leq w_{01}, w_{02} \leq 0.8$, which corresponds to the range $1.79 \leq \chi_{01}, \chi_{02} \leq 2.2$. The range of $w_{01}$ and $w_{02}$ has been deliberately chosen in order to match the values that would apply to most systems, such as PS, PVME, PE. We have verified that the phase transition we study is a *liquid-liquid* transition and not a liquid-gas transition.

In Fig. 6, we show the critical values of $\chi_{\rm scatt}^{\rm (A)}$ (●,○), $\chi_{\rm scatt}^{\rm (C)}$ (▼,▽), and $\chi_{12}$ (■,□) at the critical points for a 100/100 (filled symbols) and a 100/1000 (empty symbols) blend, respectively. *We wish to emphasize here that every data point in* Fig. 6 *represents a critical point*. We find that $\chi_{\rm scatt}^{\rm (A)} \approx \chi_{\rm NR}$ at each critical point, while $\chi_{\rm scatt}^{\rm (C)}$ remains almost *constant* over the entire range of $w_{01}$. Over the same range, $\chi_{\rm NR}$ changes by about 25% (100%) for the 100/100 (100/1000) blend. Although it is not noticeable in Fig. 6, $\chi_{\rm scatt}^{\rm (C)}$ is symmetric (asymmetric) about the minimum near $w_{01}$=0.78 for the 100/100 (100/1000) blend, and changes by a very small amount (≅ 1%), which makes the behavior of $\chi_{\rm scatt}^{\rm (C)}$ at $T_{\rm c}$ very different from its behavior away from the critical point. (In the latter case, $\chi_{\rm scatt}^{\rm (C)}$ increases with $\chi_{12}$ for all systems we have studied). The result $\chi_{\rm scatt}^{\rm (A)} \approx \chi_{\rm NR}$ should not be a surprise because for these systems $\chi_{12}$ at the critical point are large so that the blends are strongly interacting. Although we have not studied blends that have a small $\chi_{12}$ at the critical point, (weakly interacting) we conjecture that for those blends $\overline{\chi}_{\rm scatt}^{\rm (A)}$ will be a better measure of $\chi_{\rm NR}$, just as we have seen in weakly interacting asymmetric blends away from $T_{\rm c}$.

To see why $\chi_{\rm scatt}^{\rm (C)}$ remains almost a constant at the critical point irrespective of $\chi_{01}$, $\chi_{02}$ and $\chi_{12}$, we observe that $\Gamma_{\rm C}$=0 at $T_{\rm c}$. Hence,

$$\chi_{\rm scatt}^{\rm (C)} = \Gamma_{\rm C,ath} \text{ at } T_{\rm c}. \tag{29}$$

For the 100/100 blend, we note that the critical point will lie near $y$=1/2. Near $y$=1/2, $\Gamma_{\rm C,ath}$ is almost a constant [see Eq. (27b)]. With different choices of $w_{01}$, the value of $y$ at the critical point changes; however, this change is quite small. For the 100/1000 blend, we consider the empty symbols in Fig. 6; the value of $y$ still lies relatively close to the mid-range and, therefore, $\Gamma_{\mathrm{C,ath}}$ and $\chi_{\mathrm{scatt}}^{(\mathrm{C})}$ remain almost a constant.

The above findings lead us to an important conclusion that near the critical point, the most obvious correlation that $\chi_{\mathrm{scatt}}^{(\mathrm{C})}$ must increase with $\chi_{12}$ is completely lost.

## VIII. Treatment of Experimental Data

In this section we apply our procedure to extract the effective chi's from SANS data on some well known polymer blends in which experimentally determined $\chi_{\mathrm{FH}}$ appears to diverge in the composition extremes. We demonstrate that our effective chi's have weak composition-dependence in the blends we have investigated.

Let $\Gamma_{\exp}$ denote the $\Gamma$ used in SANS experiments,[37] which will give an effective chi $\chi_{\exp} \equiv \Gamma_{\mathrm{exp,ath}} - \Gamma_{\exp}$ according to our subtraction procedure. Here, $\Gamma_{\mathrm{exp,ath}}$ is the athermal part of $\Gamma_{\exp}$. On the other hand, experimentalists use $\Gamma_{\mathrm{FH,ath}}$ in place of $\Gamma_{\mathrm{exp,ath}}$. Thus, they obtain $\chi_{\mathrm{FH}}$, which contains some residual athermal part because of the incomplete cancellation as discussed in Ref. 1 and here immediately following Eq. (3). We now demonstrate how $\chi_{\exp}$ can be obtained from the experimentally extracted $\chi_{\mathrm{FH}}$. We use the method proposed in Sec. V(D) of Ref. 1. We can plot $\chi_{\mathrm{FH}}$ against $1/T$ for a given *fixed* density and determine the residual athermal part by extrapolation to $1/T\to 0$. Fortunately such plots are readily available in the literature. Let $\chi_{\mathrm{FH}} = Af(1/T) + B$; where $f(1/T)$ vanishes as $1/T\to 0$, thus making $B$ the residual athermal part. Depending on whether the athermal part ($\Gamma_{\mathrm{FH,ath}}$) underestimates or overestimates the correct athermal part ($\Gamma_{\rm exp,ath}$), we will get $B$ negative or positive, respectively. It is easy to see that $\chi_{\rm exp}$ obtained by subtracting out the true athermal part from $\Gamma_{\rm exp}$ is given by $\chi_{\rm exp} = Af(1/T) = \chi_{\rm FH} - B$. If we can show that $A$ has weak composition dependence it is clear that $\chi_{\rm exp}$ would also have a weak composition dependence. In Table II we show the value of $A$ for some well-known systems. There is a lot of uncertainty in the determination of $\chi_{\rm FH}$ for low or high $\phi_{\rm D}$ (the deuterated species density) where $I_2$ is very small. Therefore it is very difficult to determine $A$ in the composition extremes. For example, in the dPE/hPE blend (Ref. 8) when $\phi_{\rm D}$=0.044 we get $A$=0.6, which is very large from its value near the mid-range ($\phi_{\rm D}$=0.457). However it is quite remarkable that $\chi_{\rm exp}$ is almost a constant over the composition range $\phi_{\rm D}$=0.087 to 0.457. Thus, the results in Table II show that when the athermal part is cancelled *exactly*, the resulting effective chi has weak composition dependence. Note that we have extracted $\chi_{\rm exp}$ from experimental data without the intervention of *any* theory.

The term $(\Delta\bar{v})^2/2TK_T$ in Eq. (19) has been estimated[38] for polyethylene at 150°C to be in the range of $2\text{-}24\times10^{-4}$. Estimates of this term for polystyrene and polybutadiene have also been obtained.[39] From SANS data,[27] $\chi_{\rm FH}$ for a dPE/hPE blend (DP's 9196/8298) at 160°C is found to be approximately $2\times10^{-4}$ in the mid-range of composition. Thus, the last term in Eq. (19) is significant in magnitude. Therefore, $\chi_{\rm scatt}^{\rm (A)}$ can come out to be negative in weakly interacting asymmetric blends, and for such blends, we need to evaluate $\bar{\chi}_{\rm scatt}^{\rm (A)}$ to get a reliable estimate of $\chi_{\rm NR}$, as shown in Fig. 3.

For symmetric and strongly interacting asymmetric blends experimentalists only need to extract $\chi_{\text{scatt}}^{(\text{A})}$ if they wish to obtain a reliable estimate of $\chi_{\text{NR}}$.

## IX. Conclusions and Summary

In our earlier study[1] restricted to the A-ensemble, we introduced the effective chi, denoted here by $\chi_{\text{scatt}}^{(\text{A})}$, which has *no* contribution from density fluctuations. This property is also shared by an incompressible system. Thus, $\chi_{\text{scatt}}^{(\text{A})}$ shows only *minimal* compressibility-dependence, originating from the *non-randomness* in our recursive theory, and from the composition fluctuations. Thus, it appears that $\chi_{\text{scatt}}^{(\text{A})}$ might play a very useful role as a *reliable estimator* of the bare chi. Obtaining a reliable estimator is the main goal of this work. However, the A-ensemble is not a realistic ensemble. Therefore, we study here the experimentally relevant grand canonical ensemble (C-ensemble) with a special attention to their close relationship advocated in Ref. 1. The thermodynamic relationship allows us to extract quantities associated with the A-ensemble in terms of quantities associated with the C-ensemble. The prescription is independent of any particular theory. Special attention has been placed on the RMA limit of our prescription to ensure that our effective chi's either reduce to or are very close to the bare chi, when the free volume is very small.

In the RMA limit, $\chi_{\text{scatt}}^{(\text{A})}$ does not show any compressibility effect; see Eq. (25). This is not true in our non-random theory. It is found that for symmetric and strongly interacting asymmetric blends, $\chi_{\text{scatt}}^{(\text{A})} \approx \chi_{\text{NR}}$. However, for weakly interacting asymmetric blends, it can become *negative* even though $\chi_{12}$ is non-negative, showing that $\chi_{\text{scatt}}^{(\text{A})}$ does not provide a reliable estimate of $\chi_{\text{NR}}$ in weakly interacting asymmetric blends. To overcome this limitation, we introduce another effective chi with a bar. Thus, we have introduced and studied two *different* effective chi's in each of the two ensembles: the one without a bar requires an ARS, and the other with a bar requires an IRS. They are not only completely free of the experimental setup, but they also have *no* unphysical divergence anywhere. We now list important observations, some of which are also given in Table I.

We prove that $\chi_{\text{scatt}}^{(\text{C})} \geq \chi_{\text{scatt}}^{(\text{A})}$ for a blend with equal DP's, even though we have found this to be also true in all cases; see Eqs. (18,19). Their difference increases with $\Delta w$ for a given $\chi_{12} \geq 0$. Such an asymmetry-effect is also present in the RMA limit; see Eqs. (23). The bar analogs $\overline{\chi}_{\text{scatt}}^{(\alpha)}$, $\alpha$=A, or C are supposed to be better estimators of the bare chi than the unbar analogs. We have confirmed these properties in our recursive lattice theory. We have also contrasted these properties with those in the RMA limit; the latter is relevant for the F-H theory or the lattice fluid theory.[40]

### A. Symmetric Blends

For symmetric blends, we always find that $\chi_{\text{scatt}}^{(\text{C})} \cong \chi_{\text{scatt}}^{(\text{A})} \cong \chi_{\text{NR}}$, (see Fig. 1), regardless of whether we consider strongly interacting [Figs. 1 and 2], or weakly interacting (not shown here) blends. The difference, though very small, increases towards the composition extremes. We find that $\chi_{\text{scatt}}^{(\text{C})}$ decreases with increasing $\phi_0$ (see Fig. 2), or the DP ($M$). The latter behavior is similar to the behavior of the effective chi observed in incompressible blends,[18] and is not surprising.

Unfortunately, most experimental systems including isotopic blends do *not* correspond to a symmetric blend, as there is always a non-zero $\Delta\chi_0$, either due to deuteration, structural differences or due to different species. Therefore, the results obtained for symmetric blends, though highly revealing and helpful in gaining insight, are not useful for experiments.

### B. Asymmetric Blends

The results from our theory are shown in Figs. 3 and 4. The compressibility plays a dominant role now, which makes the second term in Eqs. (18,19) significant, and we find that $\chi_{\rm scatt}^{\rm (C)} > \chi_{\rm scatt}^{\rm (A)}$. Strongly and weakly interacting blends behave differently. Therefore, we discuss them separately.

#### 1. Strongly interacting Blends

The second term on the right-hand side in Eq. (14) is extremely small ($\lesssim 10^{-5}$) compared to the first term so that $\chi_{\rm scatt}^{(\alpha)} \cong \overline{\chi}_{\rm scatt}^{(\alpha)}$. They vary somewhat linearly with $y$ (for large $\Delta w$); this composition dependence comes from the asymmetry (non-zero $\delta M'$ and $\Delta\chi_0$), with $\Delta\chi_0$ having the dominant effect. The curvature of $\chi_{\rm scatt}^{(\alpha)}$ is always *positive* in all the cases that we have studied. We have found that $\chi_{\rm scatt}^{\rm (A)} \approx \chi_{\rm NR}$, but $\chi_{\rm scatt}^{\rm (C)}$ has appreciable compressibility contribution and is not a reliable estimator of $\chi_{\rm NR}$.

#### 2. Weakly interacting Blends

For weakly interacting blends, the second term in Eq. (14) is at least as important as the first term, and the values of unbar and bar chi's can be very different. We find that $\chi_{\rm scatt}^{\rm (C)}$ overestimates $\chi_{12}$ by an order of magnitude or more, and that $\chi_{\rm scatt}^{\rm (A)}$ becomes negative (see Fig. 3). The anomalous behavior is caused by the *nonzero compressibility* of the system, which can no longer be neglected. We find that $\overline{\chi}_{\rm scatt}^{\rm (C)} \cong \overline{\chi}_{\rm scatt}^{\rm (A)}$ [see Figs. 3, and 4(b)] and both predict $\chi_{\rm NR}$ almost exactly.

### C. Behavior at $T_c$

The critical $\chi_{\text{scatt}}^{(\text{C})}$ is almost a constant (with a small positive curvature) in contrast to the critical $\chi_{12}$, which can have a range of values. Thus, $\chi_{\text{scatt}}^{(\text{C})}$ completely fails to estimate $\chi_{\text{NR}}$. On the other hand, $\chi_{\text{scatt}}^{(\text{A})} \cong \chi_{\text{NR}}$, and therefore it serves as a better effective chi, even at the critical point (for strongly interacting blends).

### D. Experimental Approach

We now turn to the question of whether experiments can be done to extract the value of the bare chi's from the effective chi's. It is clear that we must *account* for compressibility effects *properly* [note the pre-factor $(1-\phi_0)^2$ in $k'_N$ and $\Gamma_C$, and the use of partial monomer volumes] in order to obtain various effective chi's. Once we know the partial monomer volumes and the compressibility; we extract $I_2$, from which $C_{12}$, and $\Gamma_C$ can be obtained. These quantities are of central interest. Note that this requires measuring the contribution $I_1$ to ensure that the right quantity is subtracted from $I(0)$. Otherwise, there will be no guarantee that $C_{12}$, and $\Gamma_C$ will not have the extraneous scattering length dependence.[31] Two general strategies that are discussed in Sec. V (D) of Ref. 1 can be used by experimentalists to extract our effective chi's. We refer the reader to Ref. 1 for details.

If measurements yield $\chi_{\text{scatt}}^{(\text{A})} \geq 10^{-3}$, we have a strongly interacting blend, for which $\chi_{\text{scatt}}^{(\text{A})} \approx \chi_{\text{NR}}$. If measurements yield $\chi_{\text{scatt}}^{(\text{A})} < 10^{-4}$, we have a weakly interacting blend. If it is a weakly interacting asymmetric blend, we need to extract $\overline{\chi}_{\text{scatt}}^{(\text{A})}$, which requires subtracting a quantity related to the IRS. However, the IRS ($\chi_{12} = 0, \chi_{01}, \chi_{02}$) does *not* exist in Nature, and the second term in Eqs. (14) *cannot* be measured. Thus, we *cannot* extract $\chi_{\mathrm{NR}}$ in weakly interacting asymmetric blends, at least at present. This is quite discouraging. However, for symmetric and strongly interacting asymmetric blends, experimentalists need to extract $\chi_{\mathrm{scatt}}^{(\mathrm{A})}$ using Eq. (18,19) if they wish to obtain a reliable estimate of the bare chi.

Different investigators have concluded differently about the relationship between $\chi_{\mathrm{scatt}}^{(\mathrm{C})}$ and $\chi_{12}$. It has been suggested recently[27] that $\chi_{\mathrm{scatt}}^{(\mathrm{C})}$ for a weakly interacting asymmetric blend is almost equal to $\chi_{12}$, which contradicts the conclusions in Refs. 24 and 25, and the effect of non-randomness implied by Eq. (1). The suggestion[27] also contradicts the conclusion drawn here.

In summary, we have shown that there exist effective chi's $\chi_{\mathrm{scatt}}^{(\mathrm{C})}$ and $\chi_{\mathrm{scatt}}^{(\mathrm{A})}$, along with their bar counterparts as measures of the energetics of the system. To quantify the usefulness of various effective chi's, we use our recursive lattice theory,[16-19] which goes beyond the RMA, for computation. For symmetric blends, $\chi_{\mathrm{scatt}}^{(\mathrm{C})} \cong \chi_{\mathrm{scatt}}^{(\mathrm{A})}$, so either of the two provides a good estimate of $\chi_{\mathrm{NR}}$. For strongly interacting asymmetric blends, $\chi_{\mathrm{scatt}}^{(\mathrm{A})}$ provides a reliable estimate of $\chi_{\mathrm{NR}}$, but $\chi_{\mathrm{scatt}}^{(\mathrm{C})}$ is not very reliable. For weakly interacting asymmetric blends, $\overline{\chi}_{\mathrm{scatt}}^{(\mathrm{A})} \cong \overline{\chi}_{\mathrm{scatt}}^{(\mathrm{C})}$, so either of the two plays the role of the reliable estimators of $\chi_{\mathrm{NR}}$. However, it does not seem possible to experimentally extract the bar quantities. Our final conclusion is that the free volume and its exchange interactions ($\varepsilon_{01}, \varepsilon_{02}$) have a strong effect on the effective chi (defined with respect to the ARS) for weakly interacting asymmetric blends to the point that it can make the effective chi an unreliable estimator of the bare chi in this case. We finally conclude that $\overline{\chi}_{\text{scatt}}^{(\text{A})} \cong \chi_{\text{NR}}$, and it is the *only* useful effective chi in all cases that we have studied.

**Table I: Summary of results**

<table>
<tr><td colspan="2">General result from thermodynamics<br>$c_{\text{scatt}}^{(\text{C})} \geq c_{\text{scatt}}^{(\text{A})} \quad (M_1 = M_2)$</td></tr>
<tr><td colspan="2">General result from our theory<br>$\overline{c}_{\text{scatt}}^{(\text{A})} @ c_{\text{NR}}$</td></tr>
<tr><td>Symmetric blends<br>$c_{\text{scatt}}^{(\text{C})} \cong c_{\text{scatt}}^{(\text{A})} \cong \overline{c}_{\text{scatt}}^{(\text{C})} \cong \overline{c}_{\text{scatt}}^{(\text{A})} \cong c_{\text{NR}}$</td><td>Asymmetric blends<br>$c_{\text{scatt}}^{(\text{C})} > c_{\text{scatt}}^{(\text{A})}$<br><br>Strongly interacting<br>$c_{\text{scatt}}^{(a)} \cong \overline{c}_{\text{scatt}}^{(a)}$, $a$ =A or C<br>$\overline{c}_{\text{scatt}}^{(\text{C})} > \overline{c}_{\text{scatt}}^{(\text{A})}$<br><br>Weakly interacting<br>$c_{\text{scatt}}^{(a)} \neq \overline{c}_{\text{scatt}}^{(a)}$, $a$ =A or C<br>$\overline{c}_{\text{scatt}}^{(\text{C})} \cong \overline{c}_{\text{scatt}}^{(\text{A})}$</td></tr>
</table>

**Table II: Determination of *A* from experimental data**

| Reference | System | Density of deuterated species ($f_D$) | A (K) |
|---|---|---|---|
| 5 | PVE12 | 0.15<br>0.5 | 0.21<br>0.22 |
| | PEE12 | 0.15<br>0.5 | 0.25<br>0.27 |
| 8 | dPE/hPE<br>($M_1 = M_2$ = 4400) | 0.044<br>0.087<br>0.131<br>0.457 | 0.6<br>0.22<br>0.2<br>0.21 |
| 6 | dPS/hPS | 0.093<br>0.19<br>0.28<br>0.48 | $3\times10^{-3}$ (within error range)<br>$3\times10^{-3}$ (within error range)<br>$3.6\times10^{-3}$<br>$2.8\times10^{-3}$ |

## Figure captions

1. Effect of $w_{01}$ and $w_{02}$ on $\chi^{(C)}_{\rm scatt}$ (●, ▼) and $\chi^{(A)}_{\rm scatt}$ (○, ∇) in a symmetric and an asymmetric blend (■, □)
2. Effect of free volume on $\chi^{(C)}_{\rm scatt}$ in a symmetric blend.
3. Various effective chi's for a typical isotopic blend of same species with molecular weight symmetry (5000/5000) and asymmetry (100/5000), weakly interacting ($\chi_{12} \cong 8\times10^{-6}$) and small $\Delta w$. Here $\chi^{(C)}_{\rm scatt}$ overestimates $\chi_{12}$ and $\chi^{(A)}_{\rm scatt}$ is of the wrong sign (5000/5000 blend). However, $\overline{\chi}^{(A)}_{\rm scatt}$ and $\overline{\chi}^{(C)}_{\rm scatt}$ for both blends lie on top of each other on the scale of the graph and are almost exactly equal to $\chi_{\rm NR} = 6\times10^{-6}$.
4. Effect of free volume on (a) $\chi^{(C)}_{\rm scatt}$ (○, ∇) and $\chi^{(A)}_{\rm scatt}$ (●, ▼), and (b) $\overline{\chi}^{(C)}_{\rm scatt}$ (○, ∇) and $\overline{\chi}^{(A)}_{\rm scatt}$ (●, ▼). Here also $\chi^{(C)}_{\rm scatt}$ overestimates $\chi_{12}$. However, $\overline{\chi}^{(A)}_{\rm scatt}$ and $\overline{\chi}^{(C)}_{\rm scatt}$ for both blends are almost exactly equal to $\chi_{\rm NR} = 6\times10^{-4}$.
5. Temperature dependence of effective chi ratios for a 100/100 blend beginning with a critical point at $w_{01}$=0.77, $w_{02}$=0.76, $w_{12}$=0.99664 keeping $y$ and $\phi_0$ fixed at 0.497 and 0.05 respectively. At the critical point, T=100 is arbitrarily assigned.
6. Plot of $\chi^{(A)}_{\rm scatt}$, $\chi^{(C)}_{\rm scatt}$ and $\chi_{12}$ at each critical point as a function of the critical $w_{01}$, with $w_{02}=1.56-w_{01}$, for a blend with DP symmetry (100/100) and DP asymmetry (100/1000). The critical $\chi^{(C)}_{\rm scatt}$ is almost a constant and does not behave like the critical $\chi_{12}$, whereas the critical $\chi^{(A)}_{\rm scatt}$ behaves the same way as the critical $\chi_{12}$, and is almost identical to the corresponding critical $\chi_{\rm NR}$.

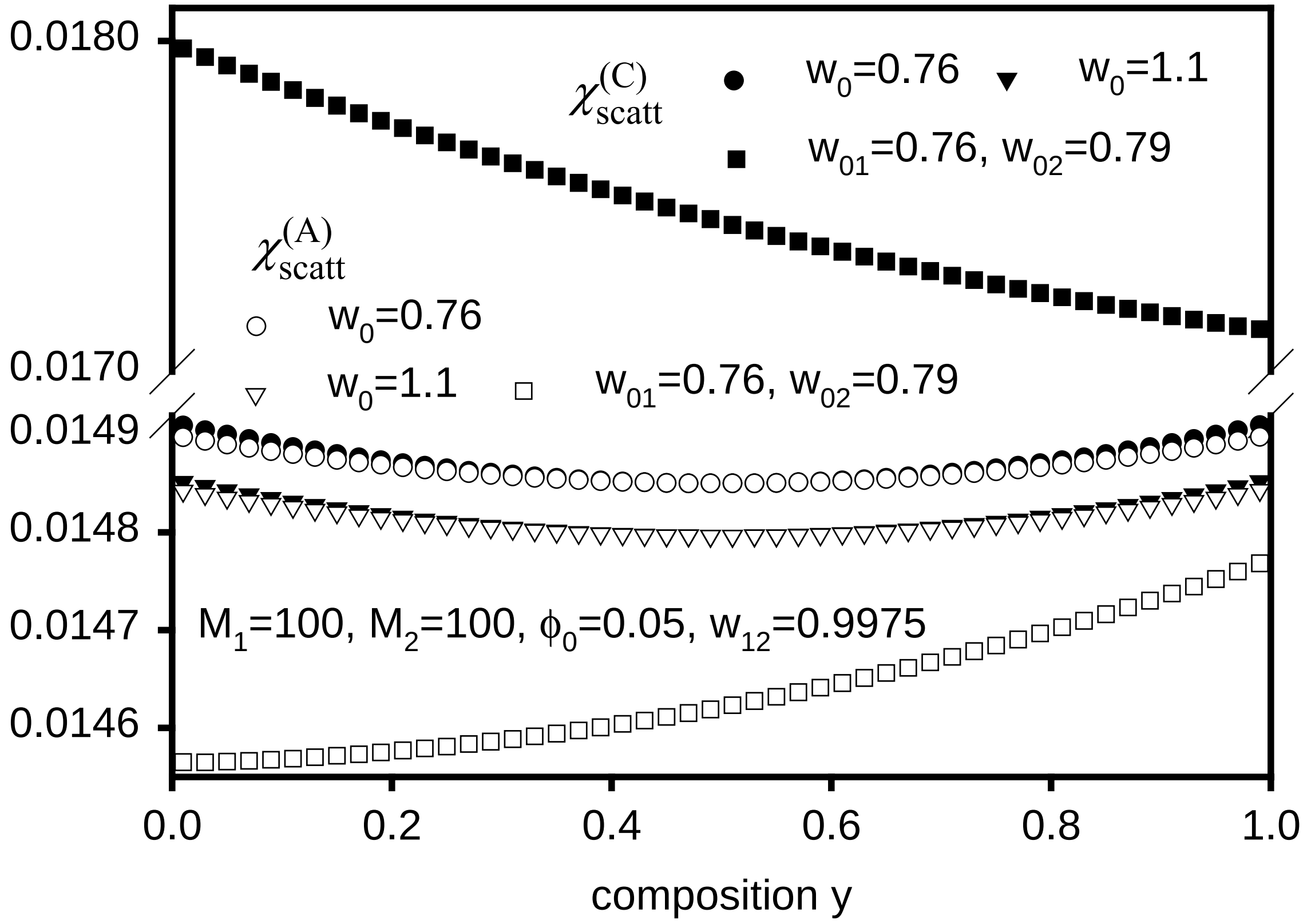

Figure 1

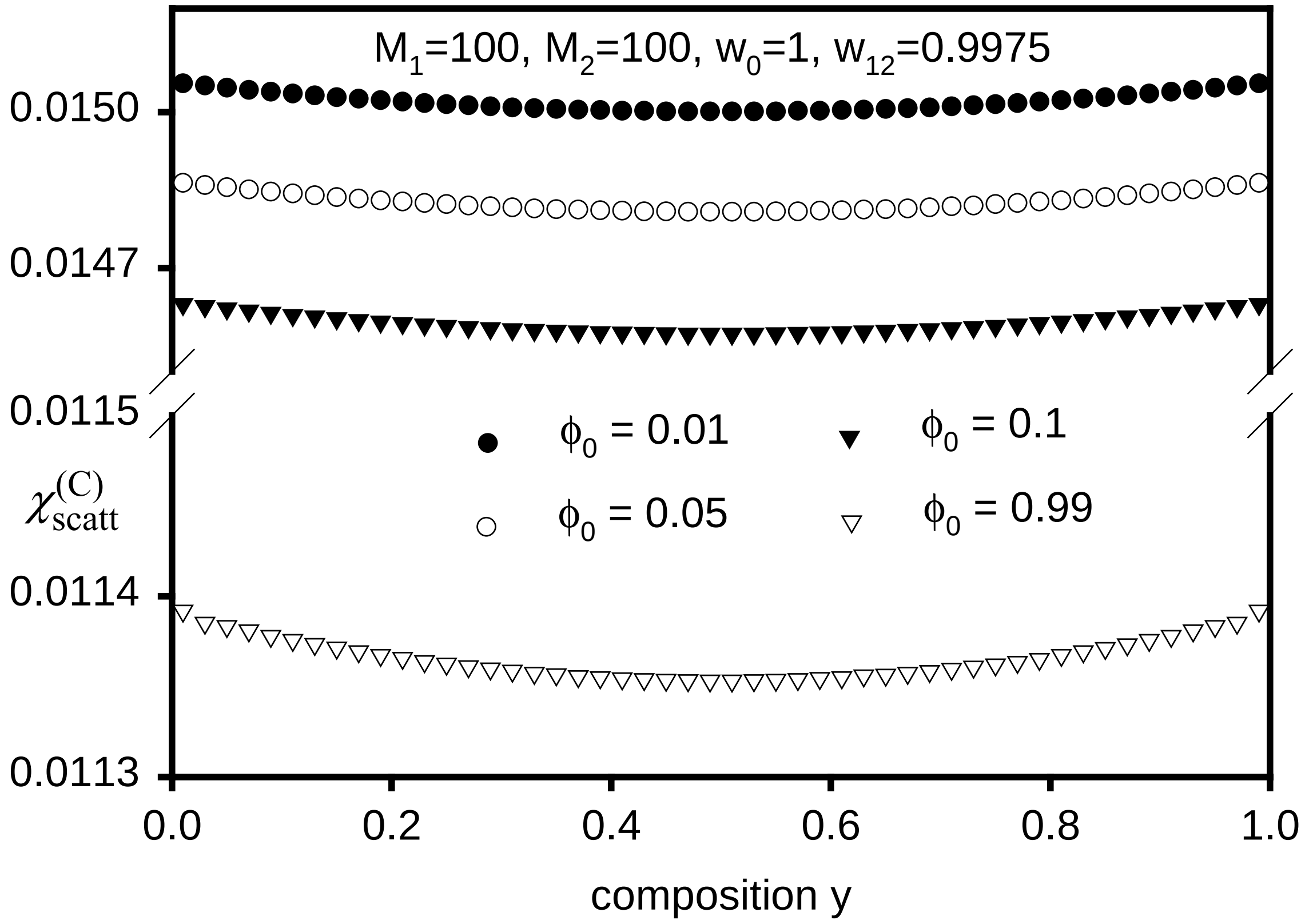

Figure 2

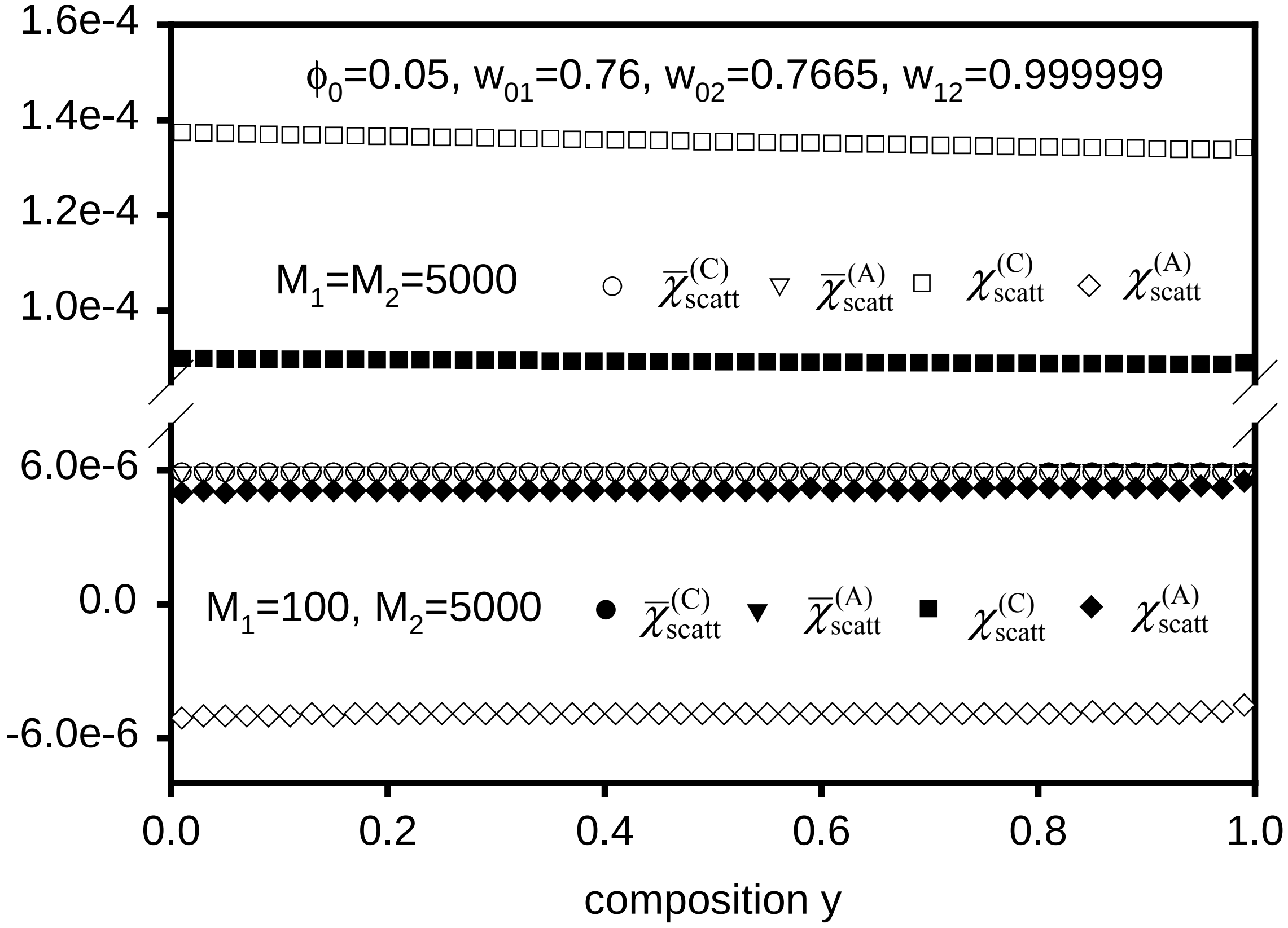

1.6e-4
1.4e-4
1.2e-4
1.0e-4
6.0e-6
0.0
-6.0e-6
$\phi_0$=0.05, $w_{01}$=0.76, $w_{02}$=0.7665, $w_{12}$=0.999999
$M_1$=$M_2$=5000
$\overline{\chi}_{scatt}^{(C)}$ $\overline{\chi}_{scatt}^{(A)}$ $\chi_{scatt}^{(C)}$ $\chi_{scatt}^{(A)}$
$M_1$=100, $M_2$=5000
$\overline{\chi}_{scatt}^{(C)}$ $\overline{\chi}_{scatt}^{(A)}$ $\chi_{scatt}^{(C)}$ $\chi_{scatt}^{(A)}$
0.0
0.2
0.4
0.6
0.8
1.0
composition y

Figure 3

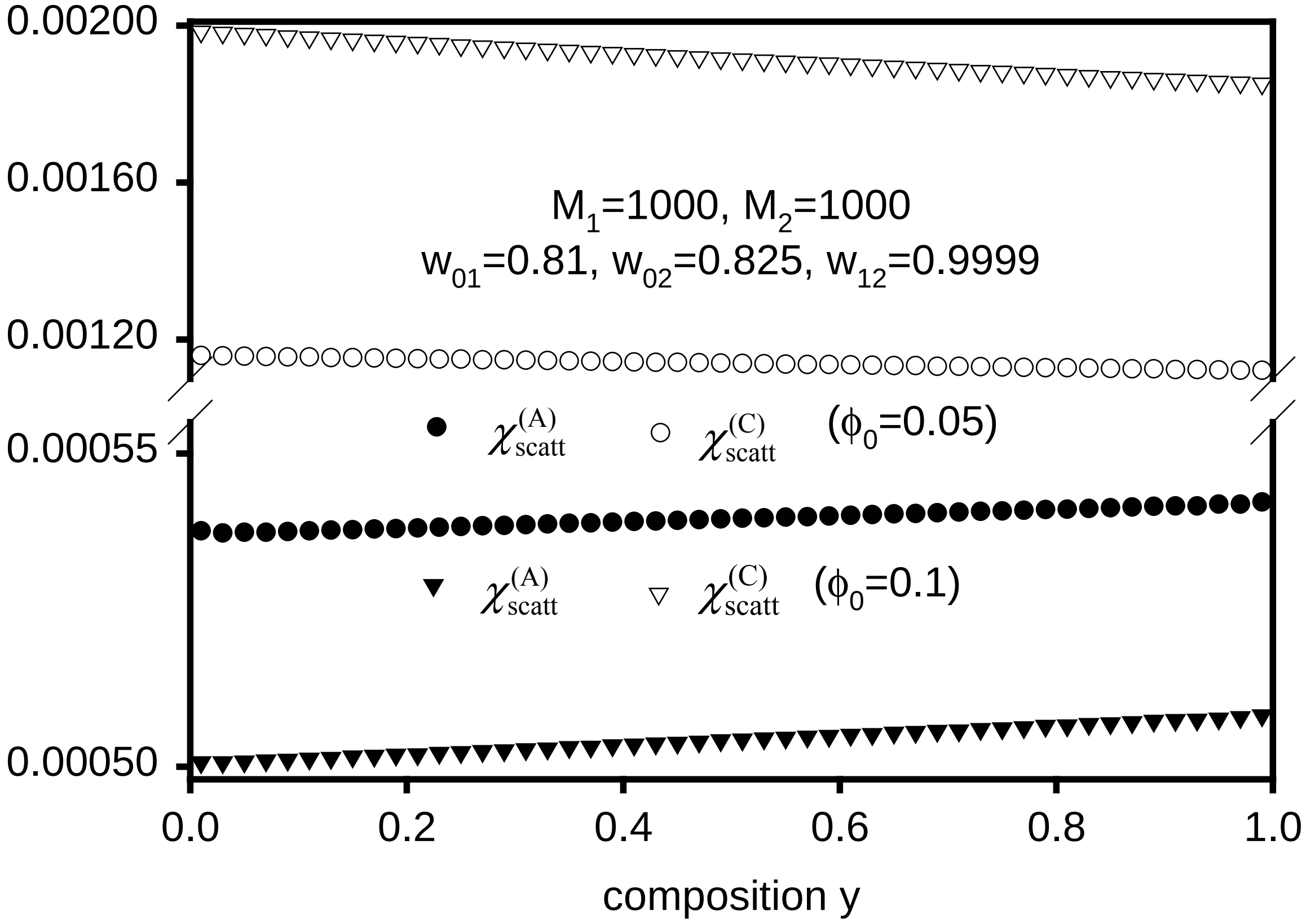

Figure 4(a)

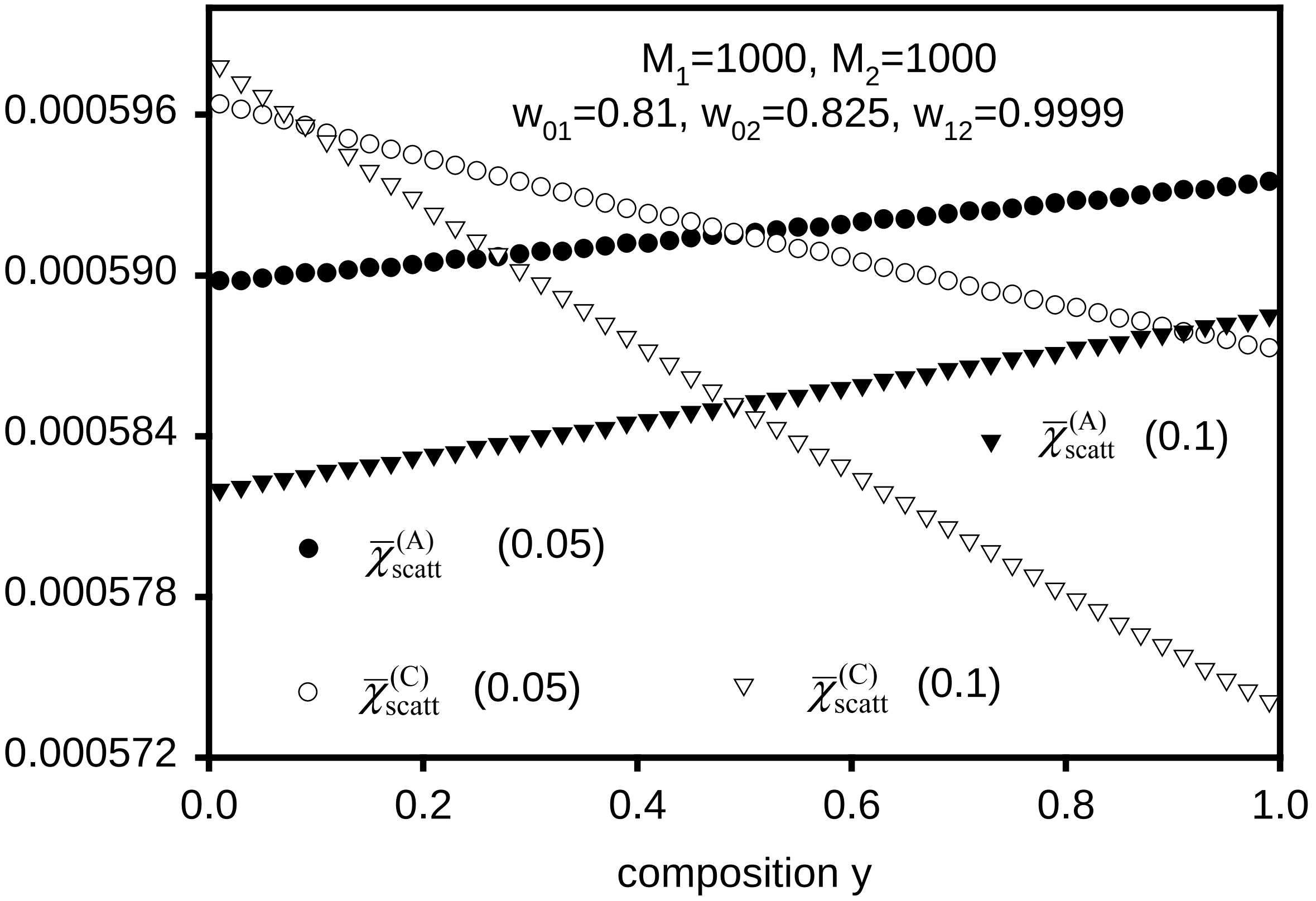

Figure 4(b)

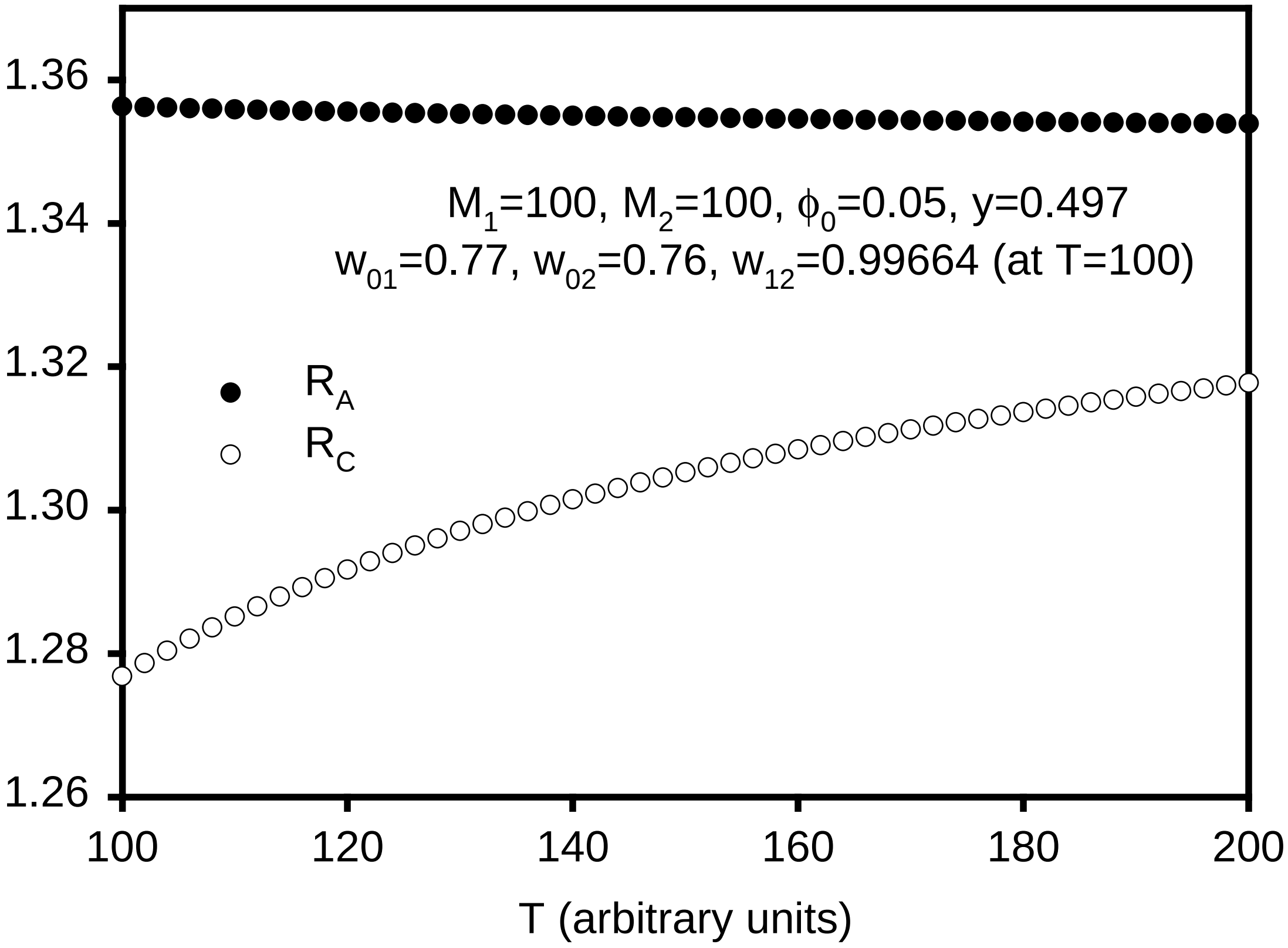

Figure 5

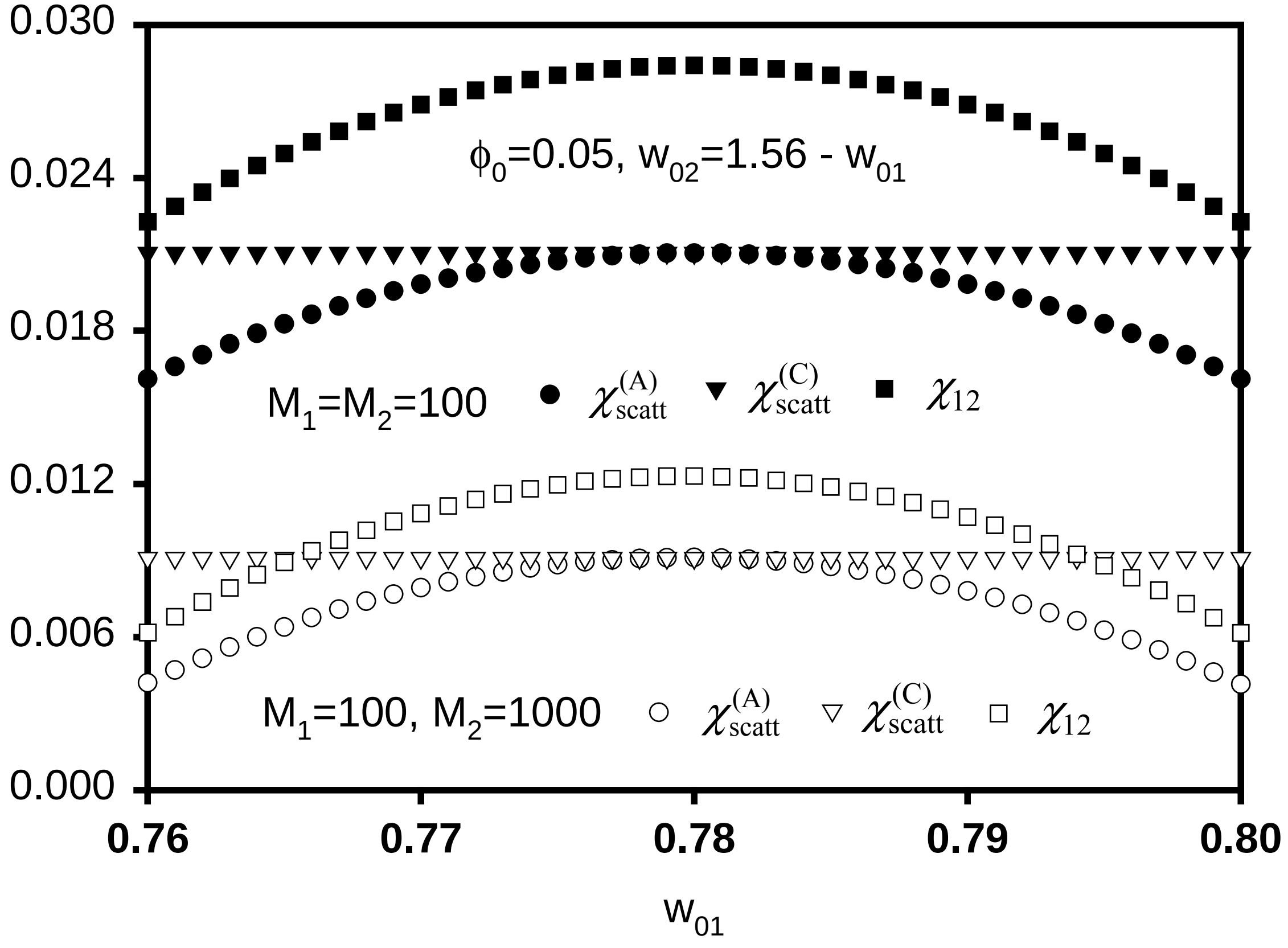
0.030
0.024
0.018
0.012
0.006
0.000
0.76
0.77
0.78
0.79
0.80
$w_{01}$
$\phi_0$=0.05, $w_{02}$=1.56 - $w_{01}$
$M_1$=$M_2$=100 $\chi_{scatt}^{(A)}$ $\chi_{scatt}^{(C)}$ $\chi_{12}$
$M_1$=100, $M_2$=1000 $\chi_{scatt}^{(A)}$ $\chi_{scatt}^{(C)}$ $\chi_{12}$

Figure 6